\documentclass[manuscript,endfloat]{geophysics}
\usepackage{float}
\usepackage{amsmath,amsfonts,amsthm}
\usepackage{graphicx}
\usepackage{multicol}
\usepackage{sidecap}
\usepackage{subfigure}
\usepackage{color}
\usepackage{lipsum}
\usepackage{url}
\usepackage{makecell}

\begin{document}

\title{Machine learning-based porosity estimation from \\ spectral decomposed seismic data}
\renewcommand{\thefootnote}{\fnsymbol{footnote}}

\address{
\footnotemark[1]University of Texas at Austin, \\ 200 E Dean Keeton St., Austin, TX 78712, USA.\\
Presently at BP plc. \\  501 Westlake Park Blvd., Houston, TX 77079, USA. \\ 
\vspace{3mm}
\footnotemark[2]Shell International Exploration \& Production Inc., \\ 150 N Dairy Ashford Rd., Houston, TX 77079, USA. \\ \vspace{3mm}
\footnotemark[3]University of Texas at Austin, \\ 200 E Dean Keeton St., Austin, TX 78712, USA.\\ \vspace{3mm}
\footnotemark[4]Lawrence Livermore National Laboratory, \\ 7000 East Ave., Livermore, CA 94550, USA.
}

\author{Honggeun Jo\footnotemark[1], Yongchae Cho\footnotemark[2], Michael Pyrcz\footnotemark[3], Hewei Tang\footnotemark[4], Pengcheng Fu\footnotemark[4]}

\footer{}
\lefthead{}
\righthead{ResUNet++ assisted porosity estimation}

\maketitle

\begin{abstract}\pagebreak
\setcounter{page}{1}
\section{Abstract}
Estimating porosity models via seismic data is challenging due to the signal noise and insufficient resolution of seismic data. Although impedance inversion is often used by combining with well logs, several hurdles remain to retrieve sub-seismic scale porosity. As an alternative, we propose a machine learning-based workflow to convert seismic data to porosity models. A \texttt{ResUNet++} based workflow is designed to take three seismic data in different frequencies (i.e., decomposed seismic data) and estimate their corresponding porosity model. The workflow is successfully demonstrated in the 3D channelized reservoir to estimate the porosity model with more than 0.9 in R2 score for training and validating data. Moreover, the application is extended for a stress test by adding signal noise to the seismic data, and the workflow results show a robust estimation even with 5\% of noise. Another two \texttt{ResUNet++} are trained to take either the lowest or highest resolution seismic data only to estimate the porosity model, but they show under- and over-fitting results, supporting the importance of using decomposed seismic data in porosity estimation.  
\end{abstract}

\section{Introduction}

\begin{figure}[]
\centering
\subfigure[]{
  \includegraphics[height=8cm]{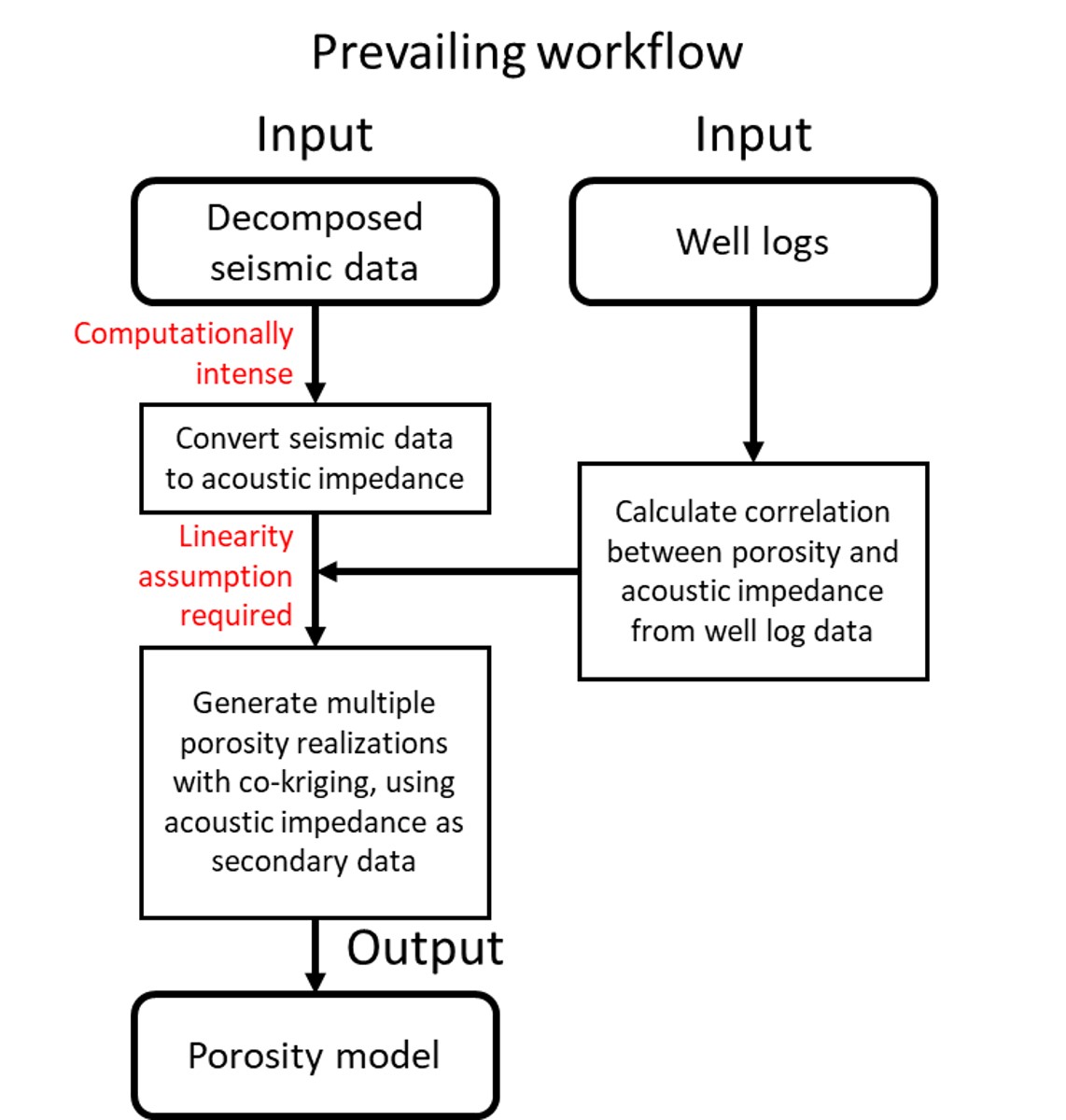}
  \label{fig:flowchart_a}}
\subfigure[]{
  \includegraphics[height=8cm]{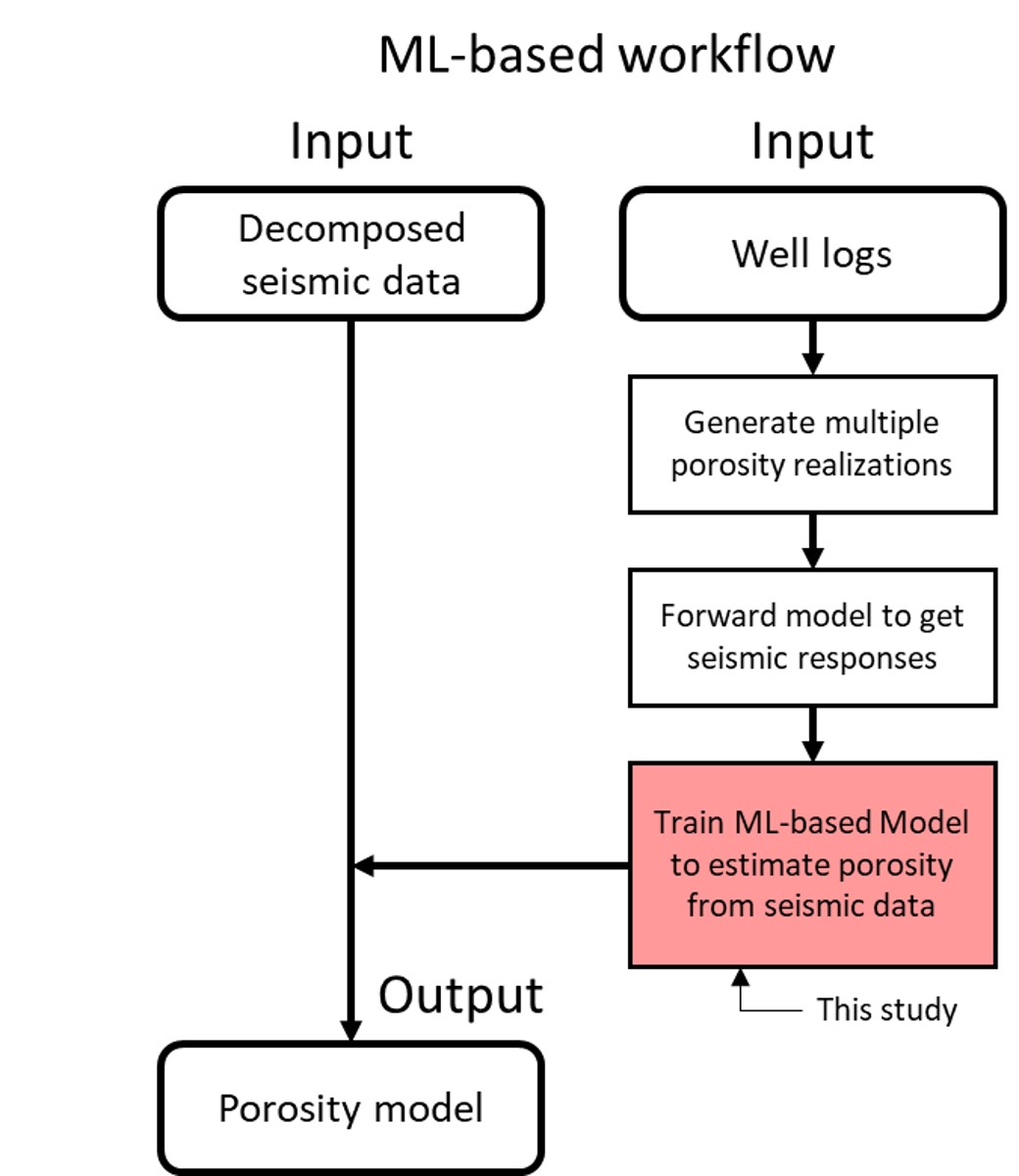}
  \label{fig:flowchart_b}}
\caption{The (a) prevailing workflow versus (b) machine learning-solution solution for porosity model estimation. This study focused on developing the ML-based model to estimate porosity from seismic data.}
\label{fig:flowchart}
\end{figure}

Integrating seismic data into subsurface modeling can help capture possible geological trends in inter-well regions \cite[]{pyrcz2014geostatistical}. In a conventional workflow as illustrated in Figure \ref{fig:flowchart_a}, (1) seismic data are first converted to a 3D acoustic impedance field, (2) the correlation between acoustic impedance and porosity is inferred from well logs (e.g., sonic log, density log, and neutron log), and (3) porosity models are constructed through co-kriging using the acoustic impedance field as secondary data, which relies on the correlation built in Step (2). 

\begin{figure}[]
\centering
  \includegraphics[width=0.8\columnwidth]{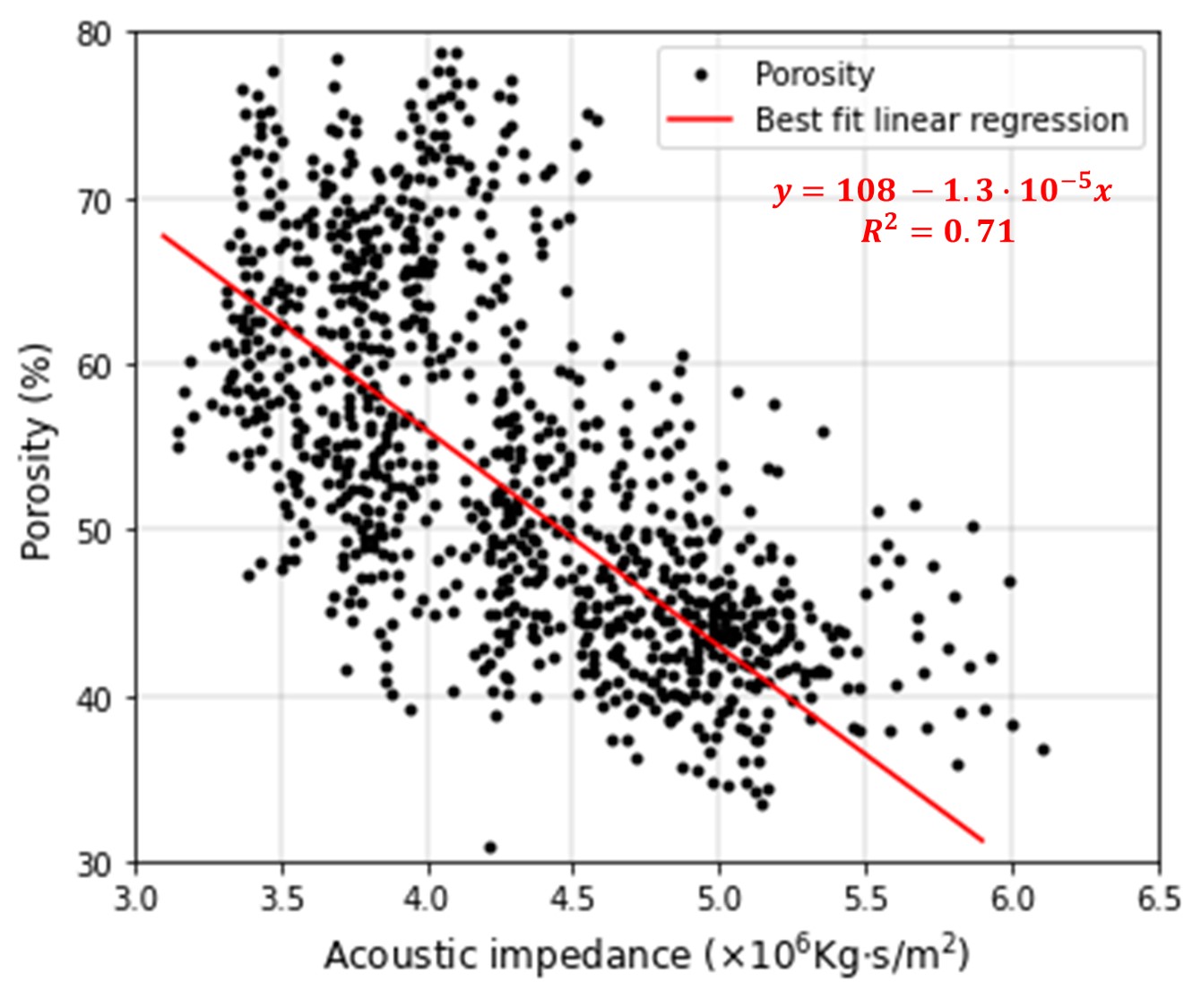}
\caption{An example of the relationship between porosity and acoustic impedance. The data is acquired from the Great Australian Bright field and the multiple well logs (e.g., sonic, density, and porosity logs) are used to collect 2,803 data points \cite[]{huuse2005seismic}. The plot shows strong heteroscedasticity, as well as non-linear relationship between porosity and acoustic impedance, making it challenge to integrate acoustic impedance by simply implementing a linear correlation.}
\label{fig:example}
\end{figure}

This method has several drawbacks. First, building the acoustic impedance model from seismic data often incurs significant computational cost, as it requires full-waveform inversion with corresponding seismic depth imaging \cite[]{cho2018quasi,cho2020accelerating}. Although employing a model-based approach can be an alternative to reduce the computing cost, this method requires manual seismic horizon picking, a very labor-intensive task susceptible to subjective interpretations \cite[]{yumin1994multiply,cho2020accelerating}. Second, this workflow relies on an overly simplistic impedance-porosity correlation established in Step (2). Typically, the correlation is assumed to be linear, and co-kriging (or co-simulation) only represents the relationship with a linear correlation coefficient \cite[]{pyrcz2014geostatistical}. However, as shown in Figure \ref{fig:example}, the actual relationship between acoustic impedance and porosity is complex and variant. Due to the poor representation of the impedance-porosity relationship, the typically large spatial connectivity of acoustic impedance often leads to overestimated porosity connectivity. In other words, sub-seismic scale heterogeneity, which is important for reservoir simulation, is not naturally honored in this workflow. Finally, most of the previous methods are not capable take advantage of spectrally decomposed seismic data. When converting seismic data to acoustic impedance, only seismic data from the highest frequency is often used, not considering the relationship among different frequency band widths. 

We hypothesize that a fully automated workflow can be built in a machine learning (ML) framework to overcome the above issues. Instead of converting seismic data to an acoustic impedance field, a properly trained ML model can convert seismic data to porosity models directly (Figure \ref{fig:flowchart_b}). The suggested ML-based workflow is composed of (1) stochastically generating a set of porosity models conditioned on well data without the use of secondary data, (2) generating spectrally decomposed seismic data from the porosity realizations using a petrophysical forward model, (3) using the data from the two above steps as “labeled data” and “features”, respectively, to train an ML model that predicts porosity fields from seismic data, and (4) feed the actual seismic data to the trained ML model to predict the porosity field. 

Out of many possible ML algorithms, we propose to apply \texttt{ResUNet++}  \cite[]{jha2019resunet++}, originally inspired by U-net \cite[]{ronneberger2015u} and ResNet \cite[]{ResNet2015} neural network structures. Although these algorithms were originally developed for image processing, such as medical image segmentation, in recent studies,
the applications are expanded to a variety of different scopes, such as fault detection from seismic sections \cite[]{li2019seismic, yang2020seismic}, enhancement of time-lapse seismic data \cite[]{jun2021repeatability}, surrogate reservoir flow simulator \cite[]{yan2020physics, maldonado2021tuning}, and proxy computational fluid dynamics in digital rocks \cite[]{santos2020poreflow, santos2021computationally}. In applying this class of ML models, we analogize 3D physical fields (e.g., velocity, porosity fields, etc.) to 3D images. In other words, “grid cells” carrying physical values are treated as “pixels” or “voxels”. \texttt{ResUNet++}’s image-to-image architecture is naturally suited for our objective as both the input, spectrally decomposed seismic data, and the output, porosity fields, attached to the same grid cells. In “training” the neural network, we are essentially establishing correlations between fields of two types of physical quantities. Inherent spatial characteristics, such as heterogeneity of porosity in shorter length scales and large spatial connectivity of acoustic impedance, are naturally “learned” by the neural network and reflected in the results, thereby overcoming a drawback of the conventional workflow. Because the training data are conditioned on well data, the trained ML model cannot be readily used to predict porosity fields at a different site. However, the conditioning of the models effectively reduce the uncertainty space, so the training data set does not need to be very large. 

In this paper, we propose an automated, ML-based porosity model estimation from the decomposed seismic data. In \textbf{Theory \& Method}, we explain (1) geostatistics for generating multiple porosity realizations, (2) geophysical forward model for converting the reservoir models to seismic responses (e.g., low-, mid-, and high-resolution seismic data), and (3) structure of \texttt{ResUNet++} to estimate the porosity model from the decomposed seismic data. In \textbf{Numerical Examples}, the suggested workflow is demonstrated in 3D channelized reservoir models and tested its robustness by (1) expanding applications to different geological scenarios and (2) adding signal noises, commonly observed in acquiring seismic data. Moreover, we perform a comparative study between using multiple-frequency seismic data and single-frequency seismic data in porosity estimation to reveal advantage in using decomposed seismic data. In \textbf{Conclusion}, we summarized observations in demonstrations and conclude our work.

\section{Theory \& Method}
\subsection{Geostatistics}
\begin{figure}[]
\centering
\subfigure[]{
  \includegraphics[width=0.47\columnwidth]{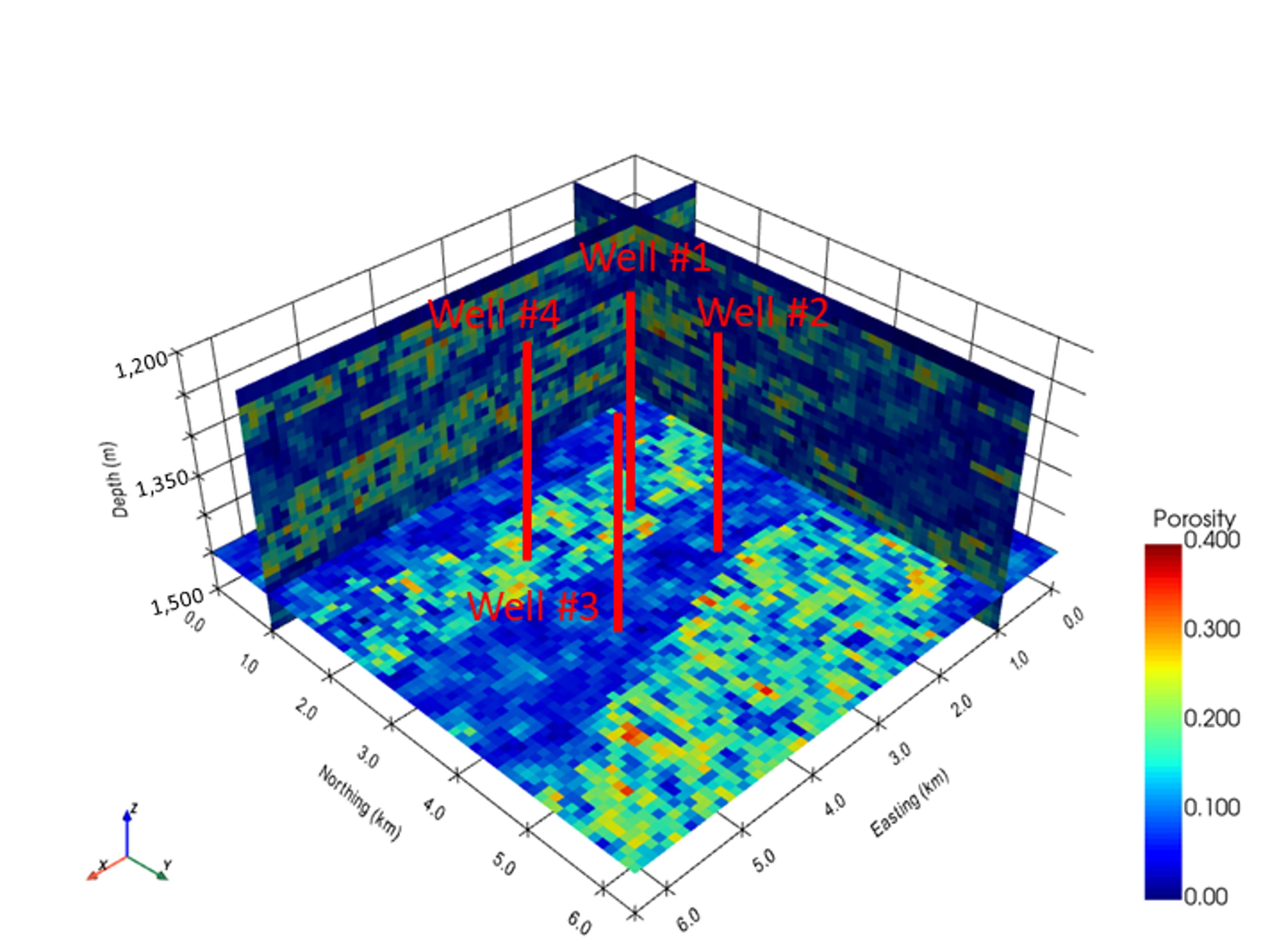}
  \label{fig:phi1}}
\subfigure[]{
  \includegraphics[width=0.47\columnwidth]{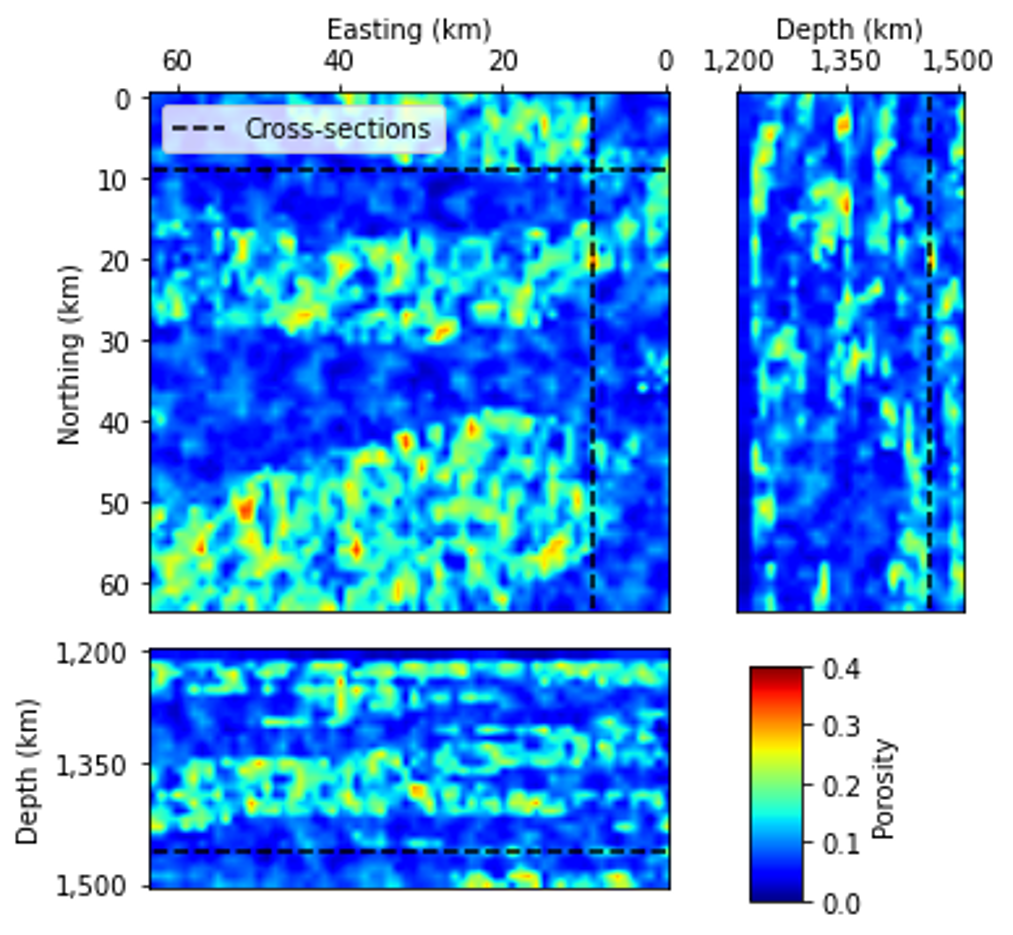}
  \label{fig:phi2}}
\subfigure[]{
  \includegraphics[width=0.47\columnwidth]{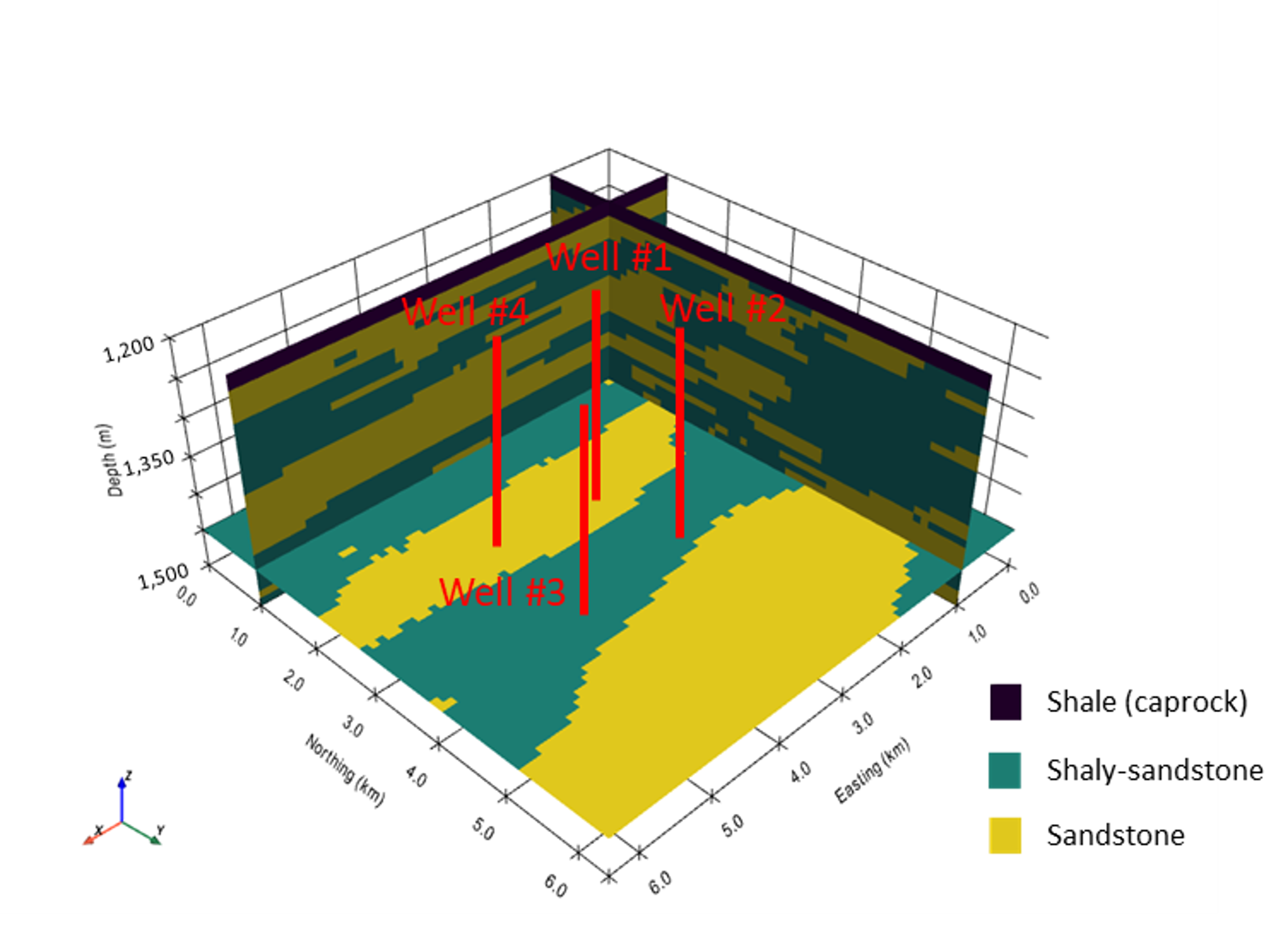}
  \label{fig:facies1}}
\subfigure[]{
  \includegraphics[width=0.47\columnwidth]{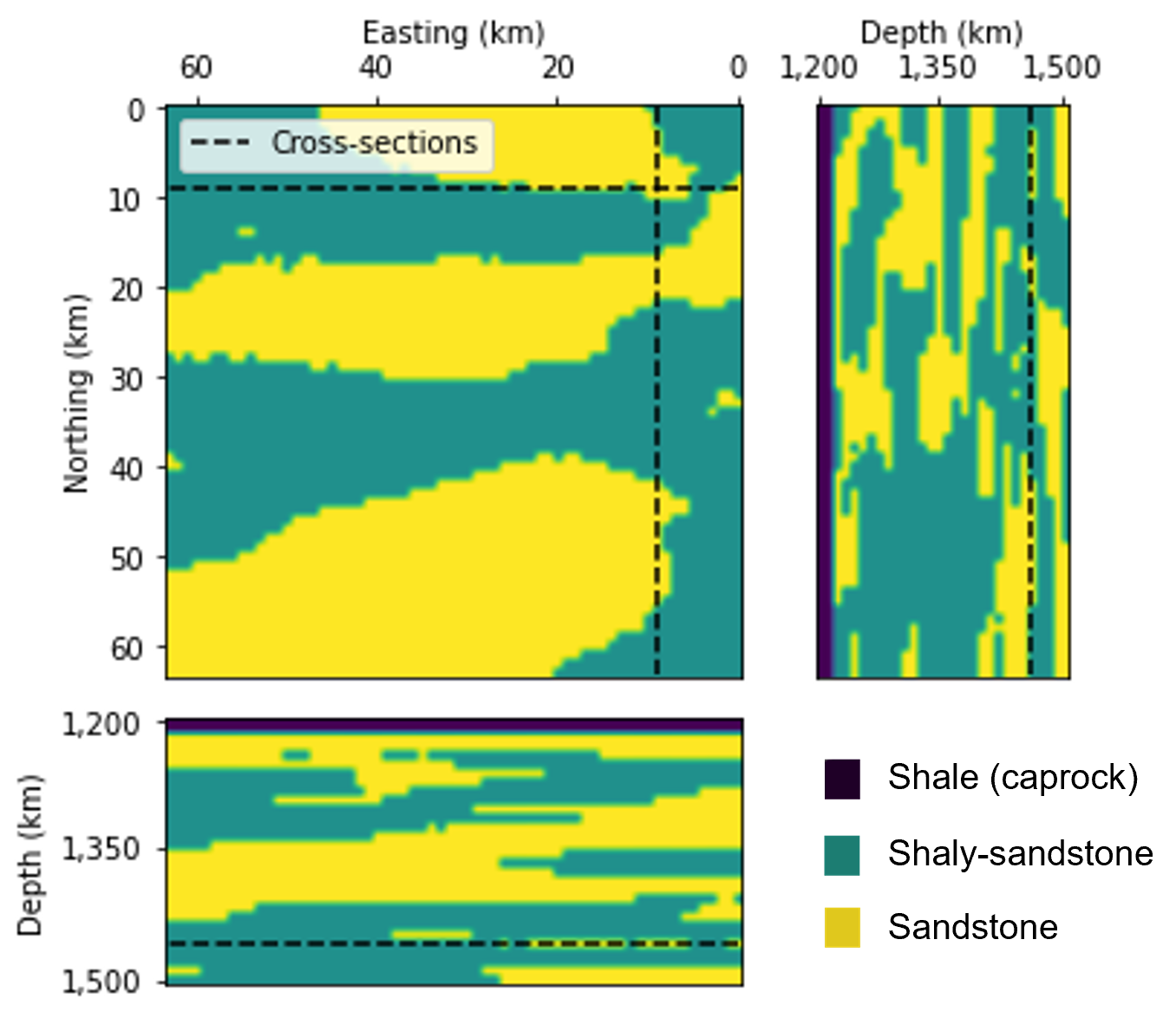}
  \label{fig:facies2}}
\caption{Porosity and rock facies models (cross- and inline-sections) in different vintages: (a) fence diagram of porosity, (b) horizontal and cross-sectional views of porosity, (c) fence diagram of rock facies, and (d) horizontal and cross-sectional views of rock facies.  The black dash-lines show the relative location of each cross- and inline-section. We applied a vertical exaggerations ($\times$100) for displaying purpose.}
\label{fig:models}
\end{figure}

The reference model (Figure \ref{fig:models}) is originally designed by \cite{bosshart2018quantifying,tang2020deep,TANG2021103488} for geologic carbon storage. The reservoir model has an aerial extent of 40.96~km$^2$ and the total thickness of 320~m, discretized into 64$\times$64$\times$32 grids. Therefore, each grid block has 100$\times$100$\times$10~m$^3$ dimensions. The first three top layers are the cap rock (e.g., shale), and the remaining layers consist of two rock facies (e.g., shaly sandstone and sandstone), as shown in Figures \ref{fig:facies1} and \ref{fig:facies2}. Sandstone has a higher porosity than shaly sandstone, and shale has the least porosity.

\begin{figure}[]
\centering
  \includegraphics[width=0.85\columnwidth]{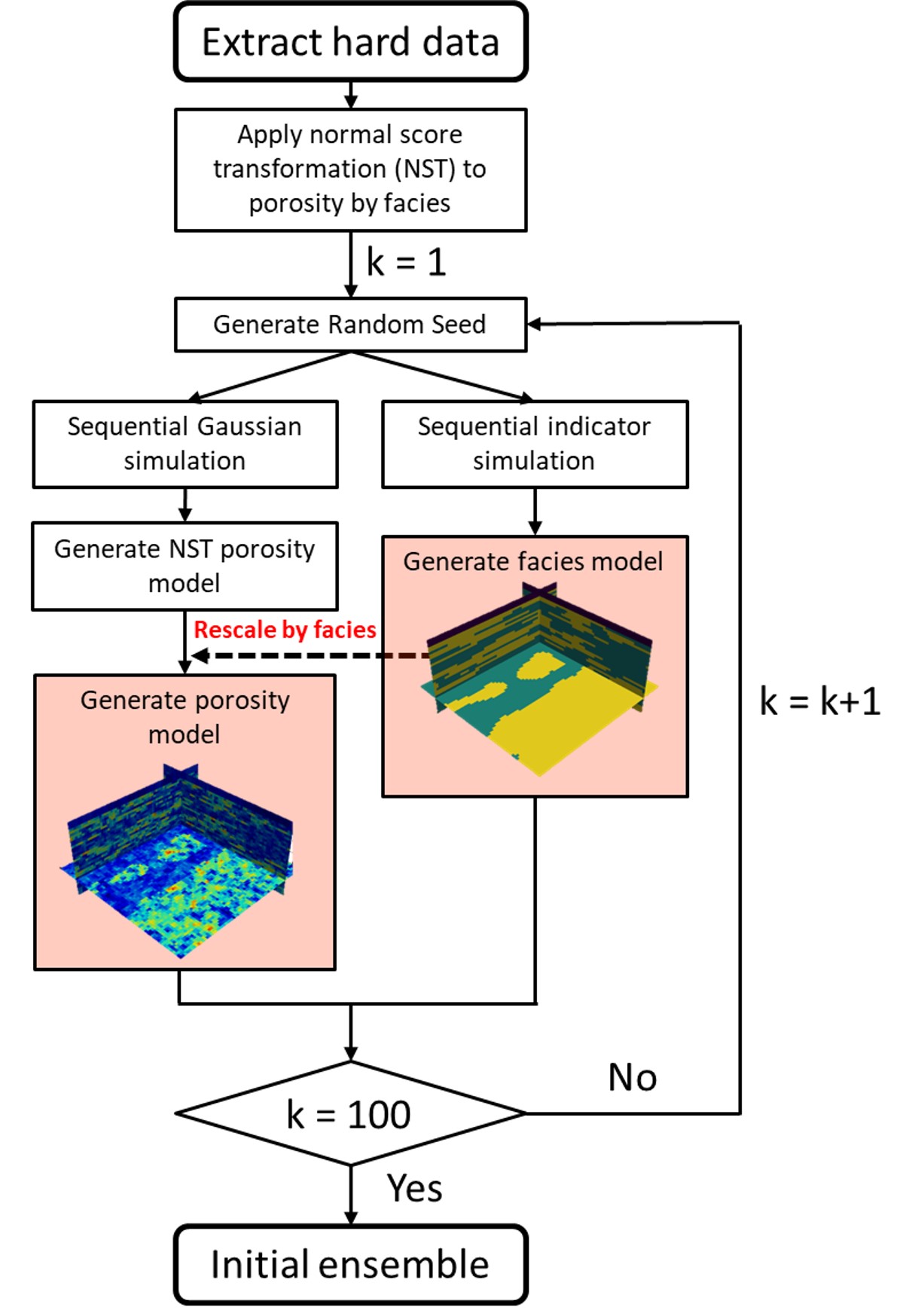}
\caption{The workflow to generate seismic data from the porosity and facies models.}
\label{fig:geostat_workflow}
\end{figure}

As ML models require a relatively large number of stochastic realizations as the training data, we generate ensemble of the reservoir models from the reference model following the workflow in Figure \ref{fig:geostat_workflow}. First, we extract the hard data (porosity and rock types) along the well trajectories (Figure \ref{fig:phi1}) from the reference model. As the reservoir model consists of more than two rock types, we applied normal-score transformation (NST) \cite[]{pyrcz2014geostatistical, jo2017history} to porosity to make all rock types have the same scale (i.e., zero means and unity standard deviation). Then, Sequential Gaussian Simulation (SGS) and Sequential Indicator Simulation (SIS) are applied to generate NST porosity and rock facies models, respectively \cite[]{pyrcz2014geostatistical}. The variogram of NST porosity is assumed to be isotropic with a horizontal connectivity range of 250~m and a vertical range of 10~m. The variogram of rock facies is assumed to be anisotropic with the 3~km and 1.5~km ranges in easting- and northing-directions. Detailed parameters of the variograms are summarized in Table \ref{table:Table}. After generating a rock facies realization, we rescale NST porosity to original porosity following their corresponding rock type.

\begin{table}
\caption{Variogram model parameters of porosity and rock facies. Porosity has isotropic variogram, while rock facies has anisotropic variogram whose azimuth is 0$^{\circ}$ (i.e., easting).} \vspace{3mm}
\label{table:Table}
\begin{tabular}{c| c| c| c| c| c| c| c}
\Xhline{3\arrayrulewidth}
Property    & Model type  & \makecell{Major \\ range} & \makecell{Minor \\ range}  & \makecell{Vertical \\ range} & Azimuth & Sill & Nugget \\
\Xhline{2\arrayrulewidth}
\makecell{Porosity \\ (SGS)}    & Exponential & 250~m       & 250~m       & 10~m &\makecell{$-$ \\ (Isotropic)} & 1    & 0      \\
\hline
\makecell{Rock facies \\ (SIS)} & Exponential & 3,000~m        & 1,500~m      & 10~m & \makecell{0$^{\circ}$ \\ (Eastward)} & 1    & 0      \\
\Xhline{3\arrayrulewidth}
\end{tabular}
\end{table}

\begin{figure}[]
\centering
  \includegraphics[width=0.99\columnwidth]{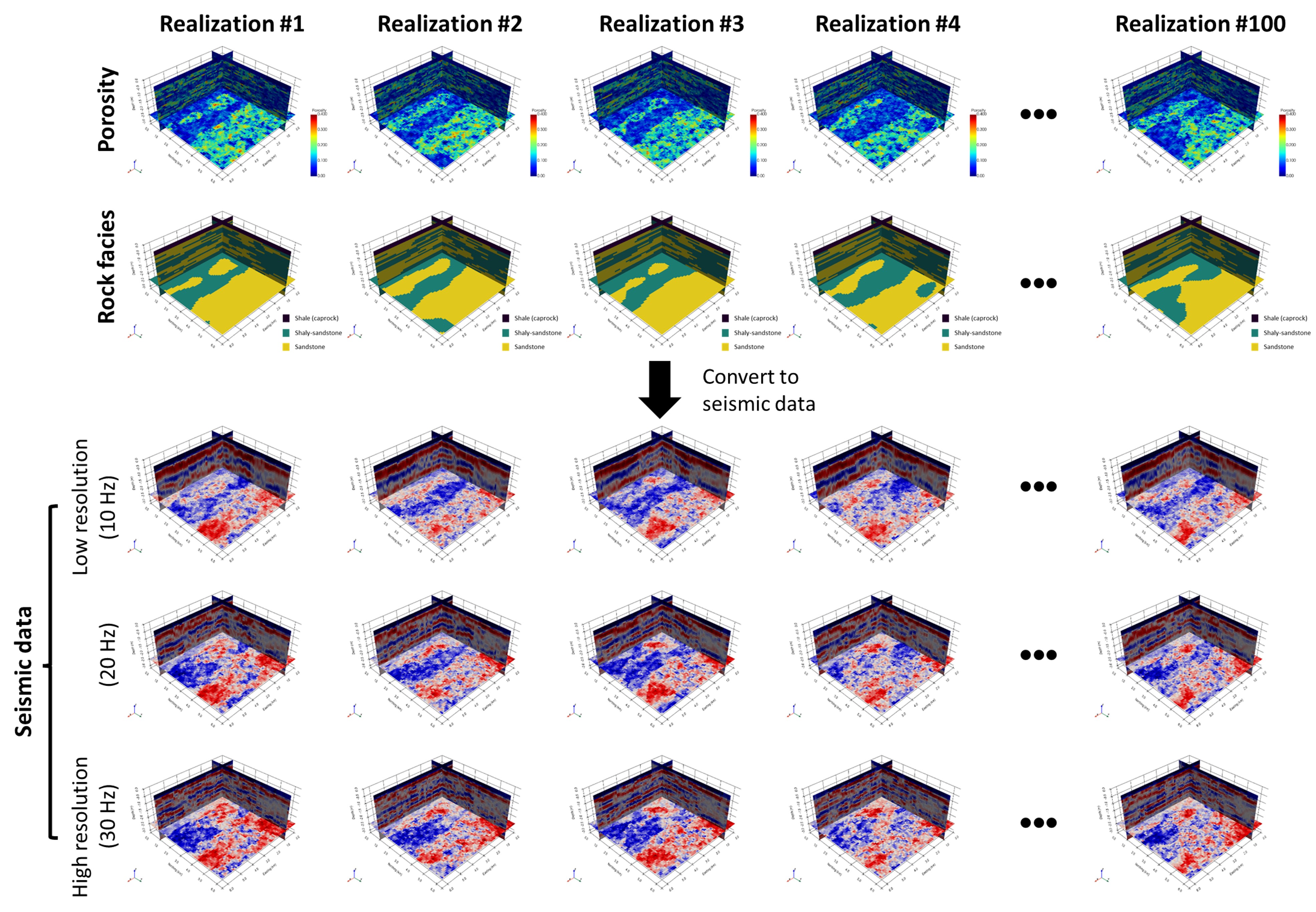}
\caption{Examples of realizations.}
\label{fig:realizations}
\end{figure}

Following the geostatistic workflow, we generate 100 realizations of porosity and rock facies models, and \texttt{GeostatsPy} \cite[]{geostat2021} is employed for SGS and SIS in this study. The first and second rows of Figure \ref{fig:realizations} presents porosity and rock facies models. Moreover, the  petrophysical model conversion is applied to generate the corresponding post-stack seismic data of realizations, as visualized in the third, fourth, and fifth rows of Figure \ref{fig:realizations}.  The three seismic data and porosity models are input and output of \texttt{ResUnet++}, respectively. 

\begin{figure}[]
\centering
  \includegraphics[width=0.99\columnwidth]{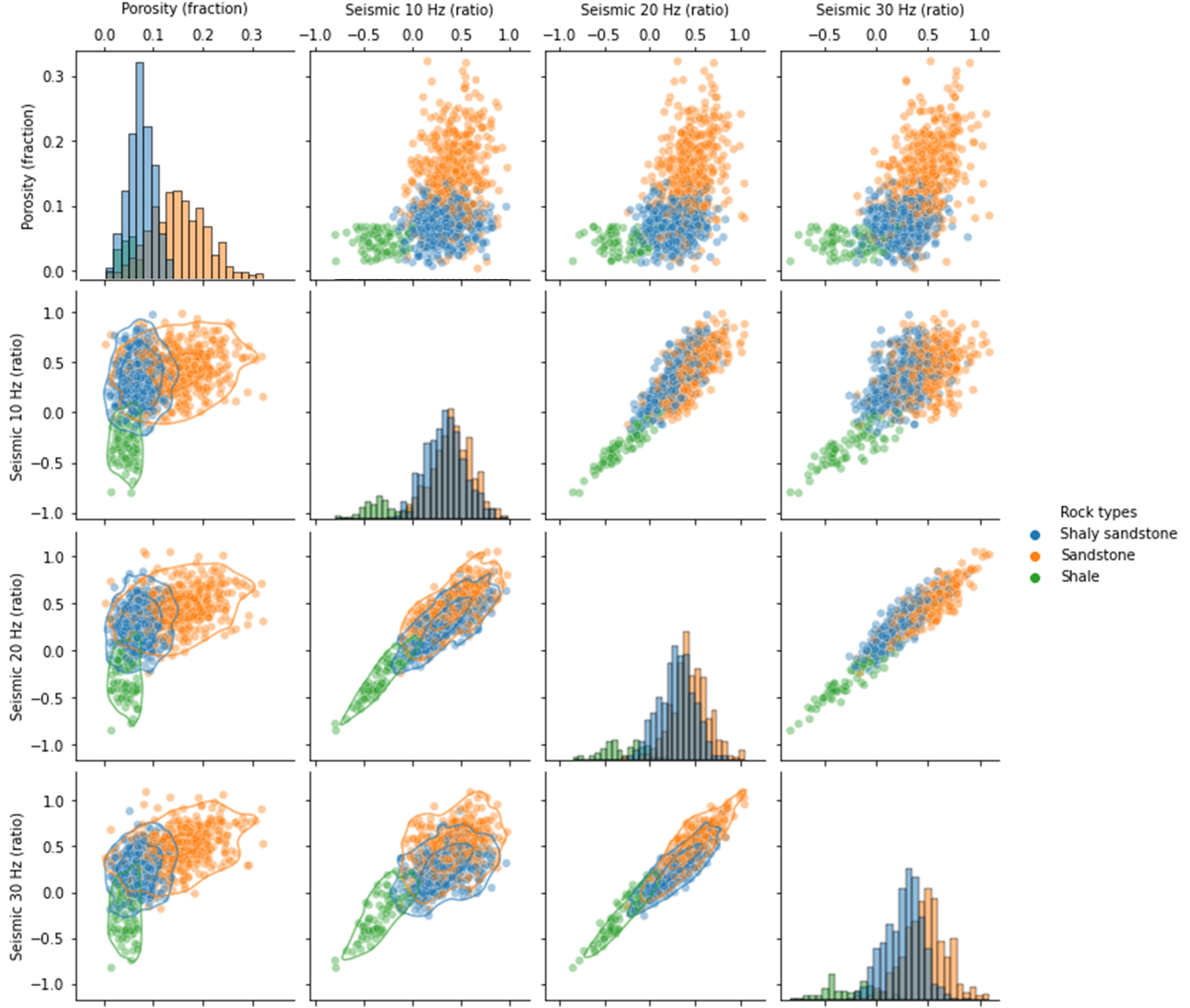}
\caption{Scatter plot matrix between porosity and seismic data with different frequencies. Blue, orange, and green points indicate shaly sandstone, sandstone, and shale, respectively. Diagonal plots are the histograms of properties, and the rest of the plots are scatter plots between two corresponding properties in the row and column.}
\label{fig:Pairplot}
\end{figure}

We perform a statistical analysis of one realization in Figure \ref{fig:realizations}. Figure \ref{fig:Pairplot} presents the relationships among porosity and three seismic data with different frequencies. Each column and row indicate different properties. Even though there is a visible porosity difference between shaly sandstone and sandstone, the difference becomes negligible in seismic data histograms, indicating seismic responses are insensitive to porosity. This is because the porosity contrast with adjacent grid blocks is more sensitive to seismic responses. Therefore, straightforward correlations between the porosity and seismic data are not available to predict porosity models, rationalizing our hypothesis of applying an ML workflow to consider these geometric features in predicting porosity models.  

\subsection{Petrophysical model conversion to seismic data}
There are a number of methods to infer the velocity from the porosity models \cite[]{wyllie1956elastic,todd1972effect,han1986effects}. In this research, we applied Eberhart-Phillips' method \cite[]{eberhart1989empirical}, in which the empirical equation is widely used for the sandstone environment. \cite{eberhart1989empirical} determined the empirical relationship (equation \ref{eq:emp}) among multiple earth parameters: velocity, effective pressure, porosity, and clay contents in sandstone core samples, which can be expressed as 
\begin{equation}\label{eq:emp}
\begin{split}
V_\text{p}^* & = 5.77 - 6.94 \phi - 1.73 \sqrt{\gamma} + 0.446 ~( P_\text{e} - e^{-16.7 P_\text{e}} ), \\
V_\text{s}   & = 3.70 - 4.94 \phi - 1.57 \sqrt{\gamma} + 0.361 ~( P_\text{e} - e^{-16.7 P_\text{e}} ),
\end{split}
\end{equation}
where $\phi$ and $\gamma$ denote porosity and clay contents, respectively. $P_\text{e}$ means effective pressure in kbar unit, and we assume that the effective pressure is linearly proportional to the depth. $V_\text{p}^*$ and $V_\text{s}$ are matrix velocity of P- and S-wave. \cite{eberhart1989empirical} applied corrections to the matrix P-velocity as follows:
\begin{equation}
\frac{1}{V_\text{p}} = \frac{1-\phi}{V_\text{p}^*} + \frac{\phi_\text{g}}{V_\text{g}} + \frac{\phi_\text{w}}{V_\text{w}} + \frac{\phi_\text{o}}{V_\text{o}},
\end{equation}\label{eq:vp_corr}
where $V_\text{p}$ is the bulk P-velocity includes the influence of fluid components. The subscript $\text{o}$, $\text{g}$, and $\text{w}$ represent oil, gas, and water, respectively. The pore space is assumed to be fully filled up with fluids ($\phi = \phi_\text{o}+\phi_\text{g}+\phi_\text{w}$), and the constant fluid velocity is applied as follows: $V_\text{o}=1300$m/s, $V_\text{g}=4800$m/s, $V_\text{w}=1500$m/s. We obtain the bulk density volume by taking average of the matrix, and fluid velocity is calculated by considering the clay contents ratio. We used the Gardner's relationship \cite[]{gardner1974formation} to make synthetic density model.

\begin{figure}[]
\centering
  \includegraphics[width=0.99\columnwidth]{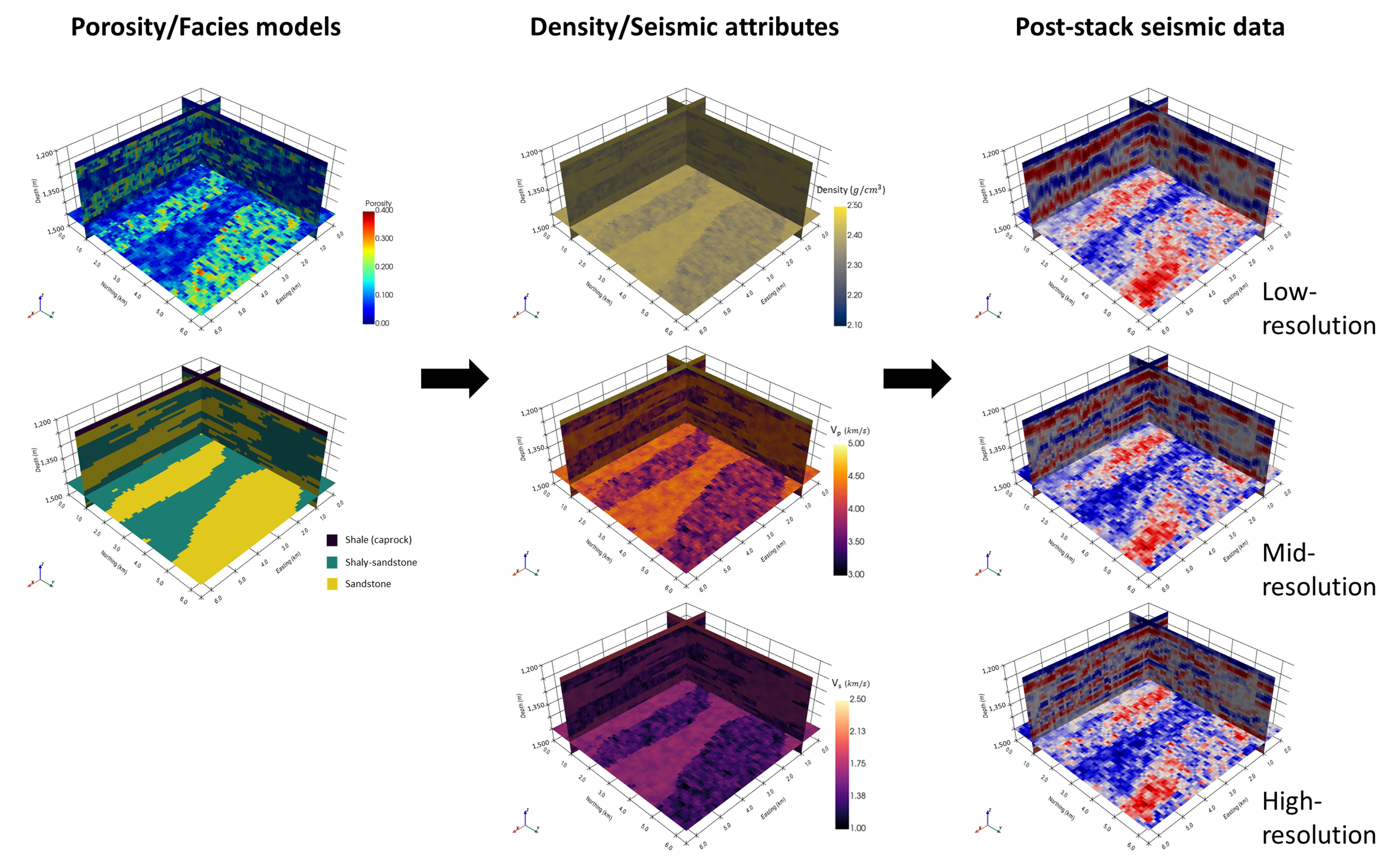}
\caption{The process to convert porosity and facies model to seismic data in multi-scales (10~Hz, 20~Hz, and 30~Hz).}
\label{fig:gen_seismic}
\end{figure}

As our suggested workflow is demonstrated to estimate the porosity models from the given post-stack seismic data, we generated synthetic seismic data by providing the porosity and rock facies models (the first column in Figure \ref{fig:models}) as an input. A schematic sketch of the workflow for seismic data generation is displayed in Figure \ref{fig:gen_seismic}. We infer the velocity and density information from the porosity and rock facies models, then calculate acoustic impedance models to convolve with the source wavelet. To demonstrate the capability of the proposed method for resolving sub-seismic scale geology, we generated three different seismic volumes with different frequency bandwidth: 10~Hz, 20~Hz, and 30~Hz, which we label high-, mid-, and low-resolution seismic data in Figure \ref{fig:gen_seismic}.

\subsection{\texttt{ResUNet++} for porosity model estimation}
\begin{figure}[]
\centering
  \includegraphics[width=0.99\columnwidth]{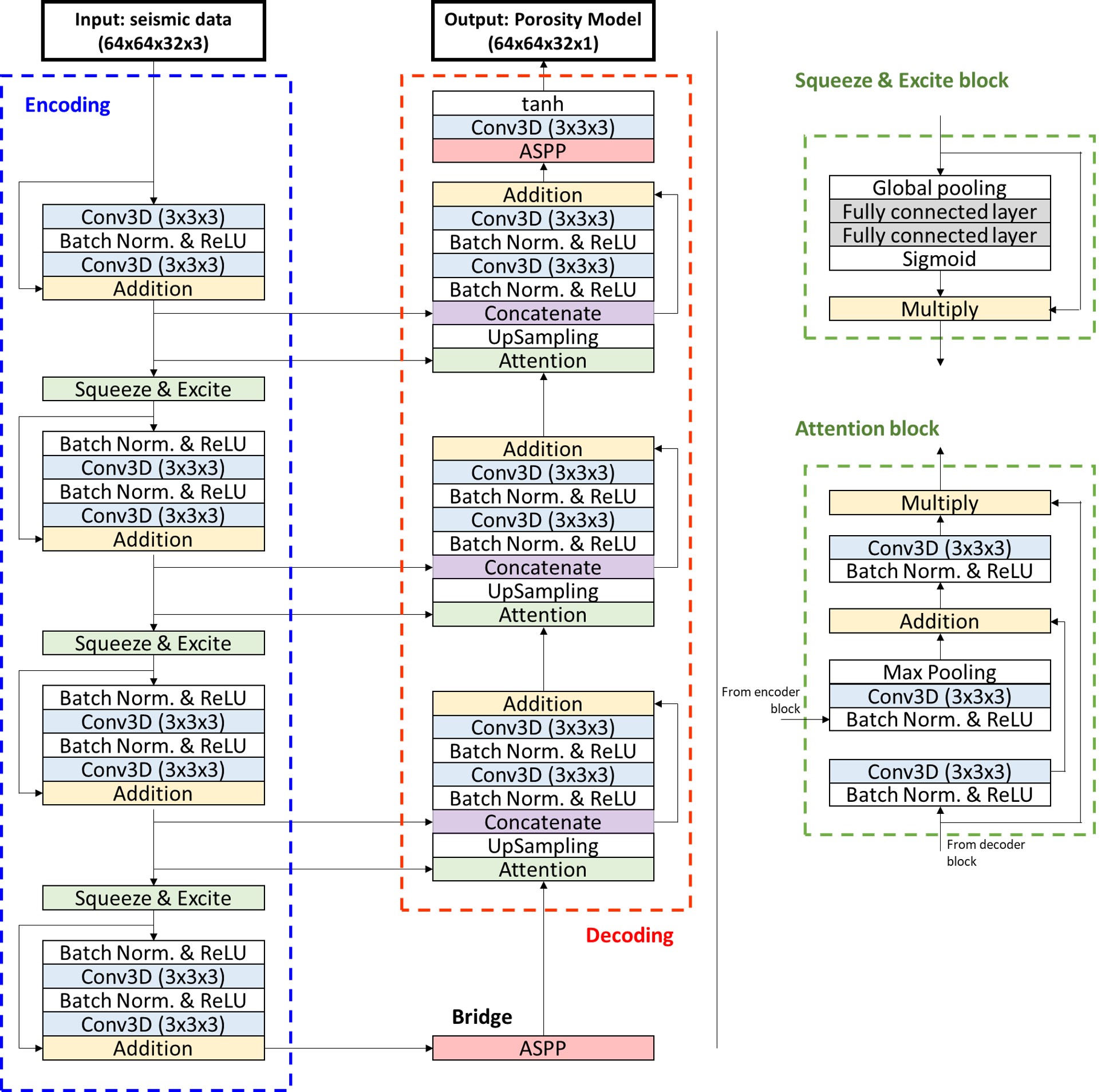}
\caption{The block diagram of the proposed \texttt{ResUnet++} architecture (modified from \cite{jha2019resunet++})}
\label{fig:ResUnet++}
\end{figure}

With a basis of convolutional neural network and ResUnet structure \cite[]{ResNet2015}, \texttt{ResUnet++} includes new advanced techniques, such as the squeeze \& excite block \cite[]{hu2018squeeze}, atrous spatial pyramid pooling (ASPP) \cite[]{chen2017deeplab}, and the attention block \cite[]{vaswani2017attention}, which enhance the stability and performance of the model. From the workflow for 2D image segmentation, we expand it for 3D image prediction (e.g., 3D seismic data and porosity models), and the architecture of neural networks is described in Figure \ref{fig:ResUnet++}. The detailed structure of \texttt{ResUnet++} is available via the author's repositories: \texttt{https://github.com/whghdrms}.    

In this study, input and output of our \texttt{ResUnet++} are seismic data (three different resolutions) and porosity model. The \texttt{ResUnet++} consists of two main steps: encoding and decoding, which are connected by a bridging block (i.e., ASPP). Through the encoding stem, the dimensions of feature maps decrease while the number of channels increases to extract main geometric features in seismic data. After passing the ASPP block, the feature maps go through the decoding stem, where their dimension is expanding back to the original reservoir size (i.e., 64$\times$64$\times$32). In restoring the original dimension, the encoded seismic information is transformed to the corresponding porosity model. 

Typical neural network operations (e.g., convolution, batching normalization, activation functions, up-sampling, and pooling) are visualized in white and blue blocks in Figure \ref{fig:ResUnet++}. \cite{jun2021repeatability} extensively explained the mathematical backgrounds of these operations. The yellow blocks indicate element-wise addition and multiplication, and the purple blocks denote concatenating feature maps over channels. Especially, the addition and concatenation are for skip connections (or shortcuts) that adjusts the complexity of the neural network flexibly \cite[]{ResNet2015, ronneberger2015u}. The green blocks are the squeeze \& excite block and attention block whose detailed sequence are visualize in the right side of Figure \ref{fig:ResUnet++}. The key idea of these two blocks is to emphasize regional information of feature maps to improve the performance of the neural network model \cite[]{vaswani2017attention, hu2018squeeze}, which are widely used for recent natural language processing studies, such as \cite{wang2018glue, young2018recent}. The red blocks indicate the ASPP, a technique for resampling features in multi-scales \citep{he2015spatial}. 

The seismic data and porosity models are re-scaled through Min-max normalization to have the range between -1 and 1, as described in the following equations:
\begin{equation}
\begin{split}
{\phi_\text{normalized}} = \frac{\phi-\phi_{\min}}{\phi_{\max}-\phi_{\min}}\times2-1, \\
{S_\text{normalized}} = \frac{S-S_{\min}}{S_{\max}-S_{\min}}\times2-1, 
\end{split}
\end{equation}\label{eq:MinMax}
where $\phi$ and $S$ are porosity and seismic data, and subscript $\text{normalized}$, $\min$, and $\max$ represent normalized (or re-scaled), minimum, and maximum values, respectively.

The mean squared error (MSE) of the estimated porosity is used as a loss function ($L$) in the proposed \texttt{ResUnet++} and can be expressed as follows:
\begin{equation}
{L} = \frac{1}{N_\text{batch} \times N_\text{grid}}
\sum_{n=1}^{N_\text{batch}} \sum_{i,j,k=1}^{N_\text{grid}} (\hat{\phi}_{n,i,j,k} - \phi_{n,i,j,k})^2 
\end{equation}\label{eq:loss}
where $N_\text{batch}$ and $N_\text{grid}$ are the number of training data in each training batch and the number of grid cells in each porosity model. Moreover, the subscript $n$, $i$, $j$, and $k$ indicate the $n$-th training data and values at the ($i$, $j$, and $k$)-th grid cells in x-, y-, and z-directions, respectively. $\hat{\phi}$ and $\phi$ are estimated and true porosity. 

\section{Numerical examples}   
\subsection{Training and validation data}
We demonstrate the \texttt{ResUnet++} assisted porosity estimation workflow with the ensemble of 3D porosity models and seismic data (Figure \ref{fig:realizations}). The seismic data from three different frequencies (e.g., 10~Hz, 20~Hz, and 30~Hz) are used as inputs, and the porosity models are the output of \texttt{ResUnet++}. Out of 100 realizations, 70 realizations are used to train the neural networks, and the rest are retained and used for validation. Min-max normalization is a prerequisite for implementing the neural network model to remove the impact of different scales of properties (e.g., porosity and seismic data). 

\subsection{Porosity estimation via \texttt{ResUNet++}}
We train \texttt{ResUnet++} for 1,000 epoch with a batch size of 25, requiring about 5 hours on a desktop computer with an environment of CPU Intel E5-1650 3.60 GHz and GPU NVIDIA Quadro M6000 24 GB. The stop criterion of training \texttt{ResUnet++} is when the loss is stabilized and shows no decrease further in both training and validating data.  For deployment, the trained \texttt{ResUnet++} predicts a porosity model in less than a second.

\begin{figure}[]
\centering
  \includegraphics[width=0.85\columnwidth]{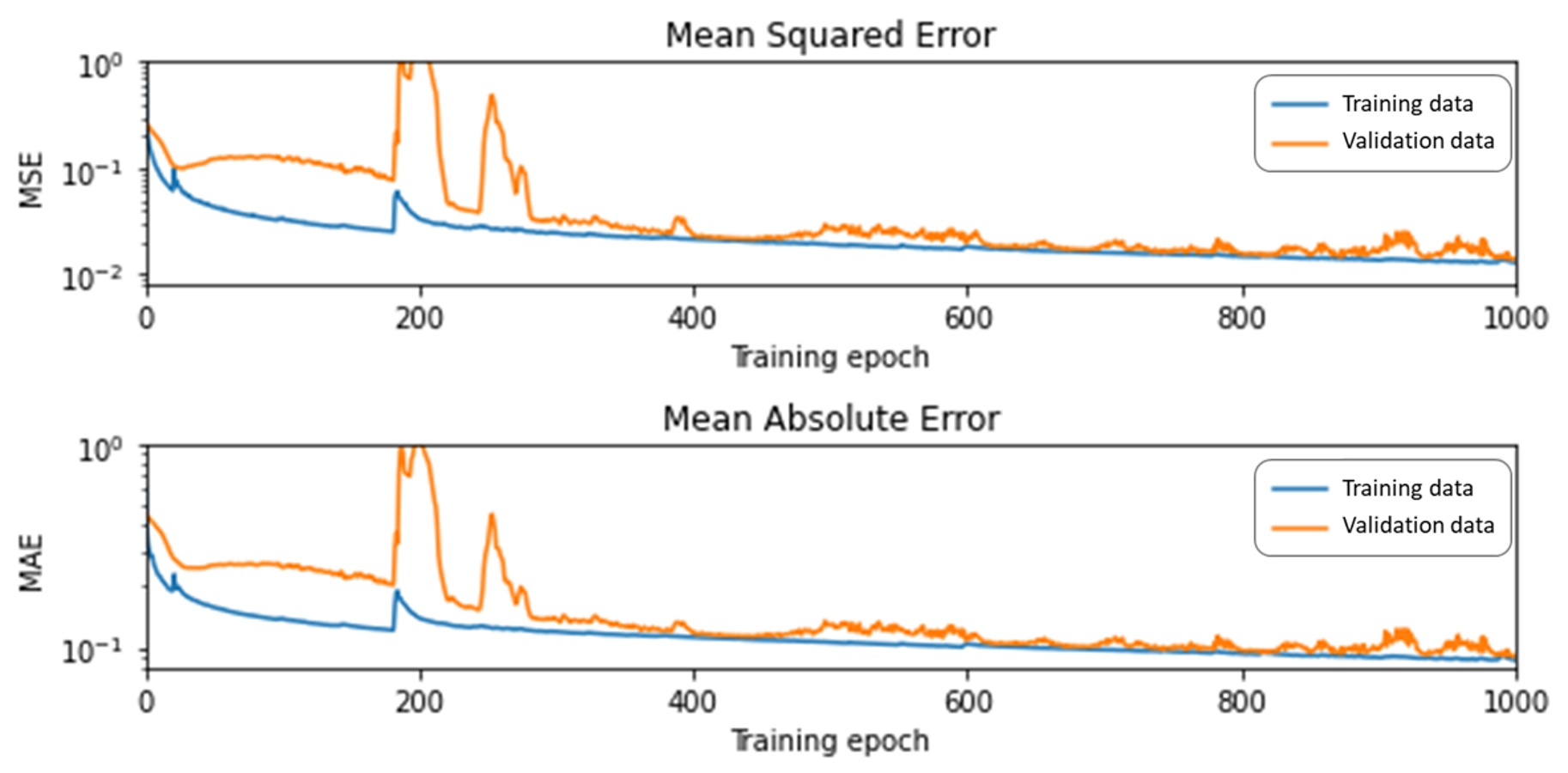}
\caption{Loss curves of the \texttt{ResUnet++} over 1,000 epochs}
\label{fig:Loss_standard}
\end{figure}

As shown in the upper plot of Figure \ref{fig:Loss_standard}, a significant discrepancy between losses of training and validating data are observed until 300 epochs, but both of them decrease simultaneously between 400 and 1,000 epochs, indicating a stable training process. Moreover, no over-fitting is observed with the given neural network structure, as the difference between training and validating loss is minimal in the end. As a secondary metric, we also compute mean absolute error (MAE), and MAE shows a similar pattern as MSE (lower plot of Figure \ref{fig:Loss_standard}). 

\begin{figure}[]
\centering
  \includegraphics[width=0.99\columnwidth]{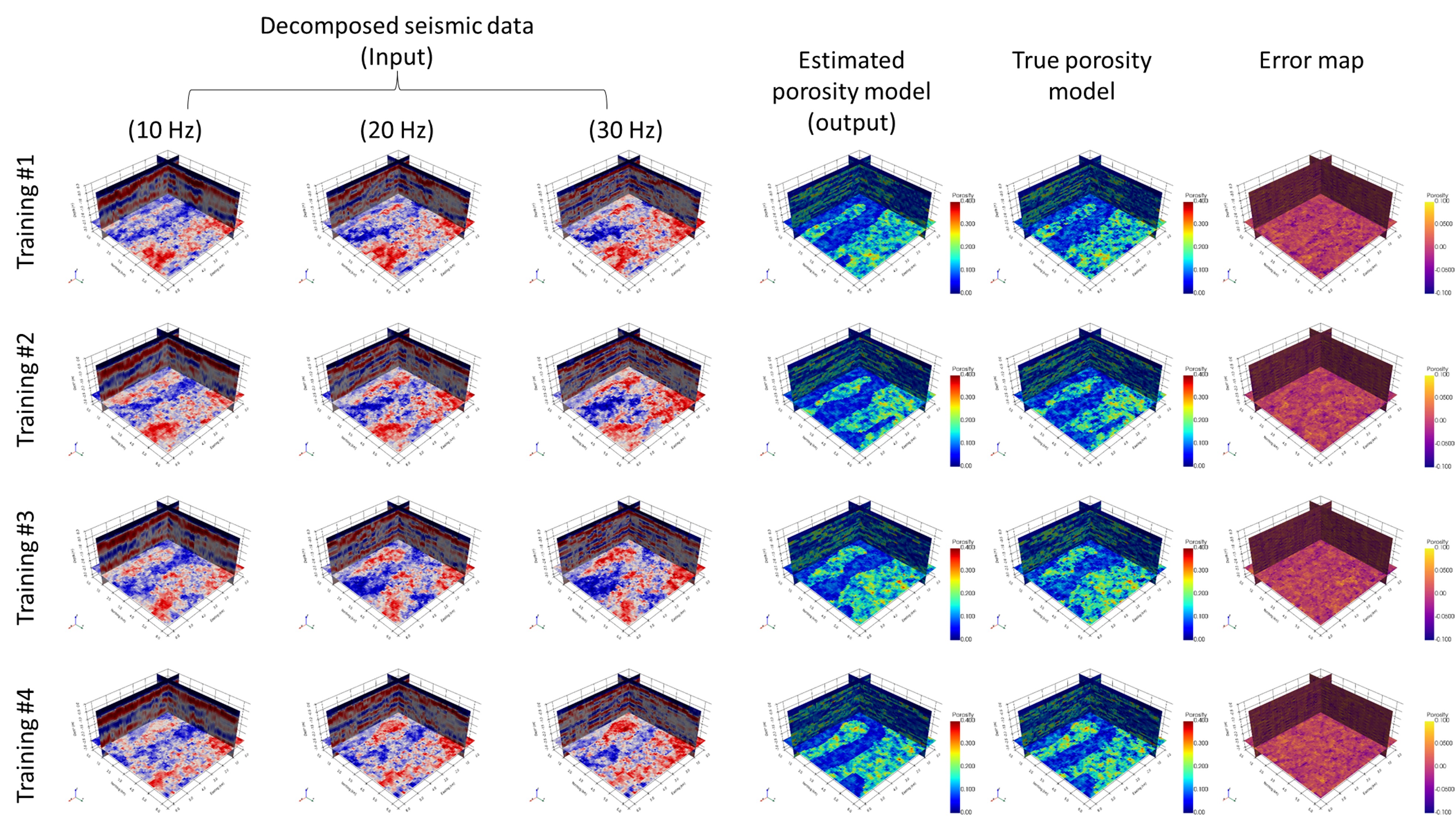}
\caption{Porosity estimation for training data}
\label{fig:demo_train}
\end{figure}

Figure \ref{fig:demo_train} presents the demonstration of porosity estimation from training data. Out of 70 training data, four realizations are visualized in each row of Figure \ref{fig:demo_train}. In the figure, the first three columns show seismic data in different resolutions, and the fourth column shows the estimate of the trained \texttt{ResUnet++}. The fifth and sixth columns depict the truth model and error between the estimate and truth, respectively. Even though the detailed heterogeneity in estimated porosity models is not identical to the truth models, the global trend is well preserved in the estimated porosity model. Furthermore, errors are close to zero and stochastically distributed in space, indicating there is no systemic bias in the trained \texttt{ResUnet++}.         

\begin{figure}[]
\centering
  \includegraphics[width=0.99\columnwidth]{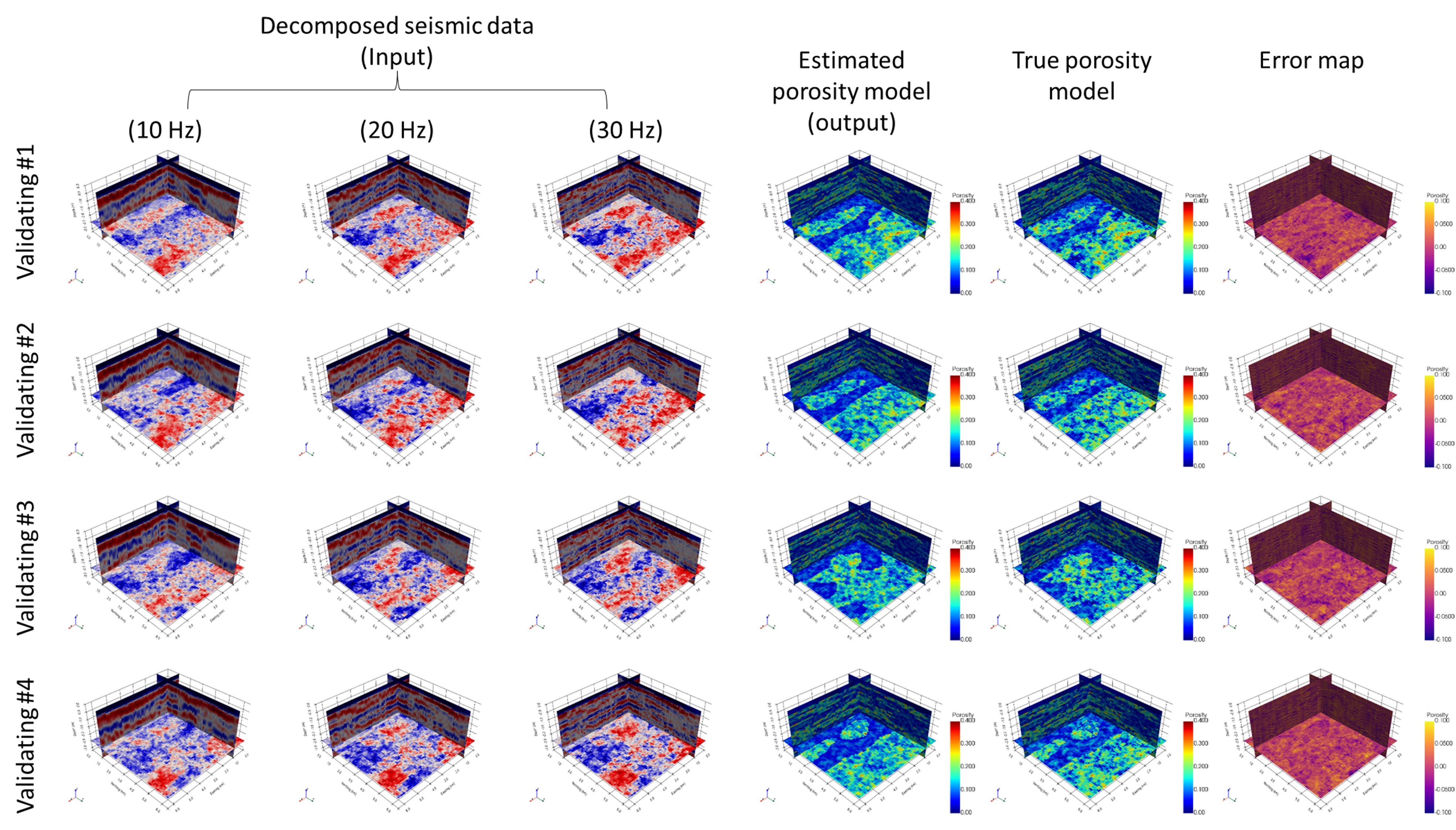}
\caption{Porosity estimation for validating data}
\label{fig:demo_validation}
\end{figure}

Figure \ref{fig:demo_validation} displays porosity estimation from four validating data. Same as the demonstration of training data, the estimated porosity models successfully capture global trends. Especially, even though the third validating data shows connected high porosity regions in the northing direction, which is not observed commonly in training data, the \texttt{ResUnet++} successfully predicts the porosity model that follows such trend. The error maps also show near zero in the entire extent, as shown in the last columns of Figure \ref{fig:demo_validation}.

\begin{figure}[]
\centering
  \includegraphics[width=0.99\columnwidth]{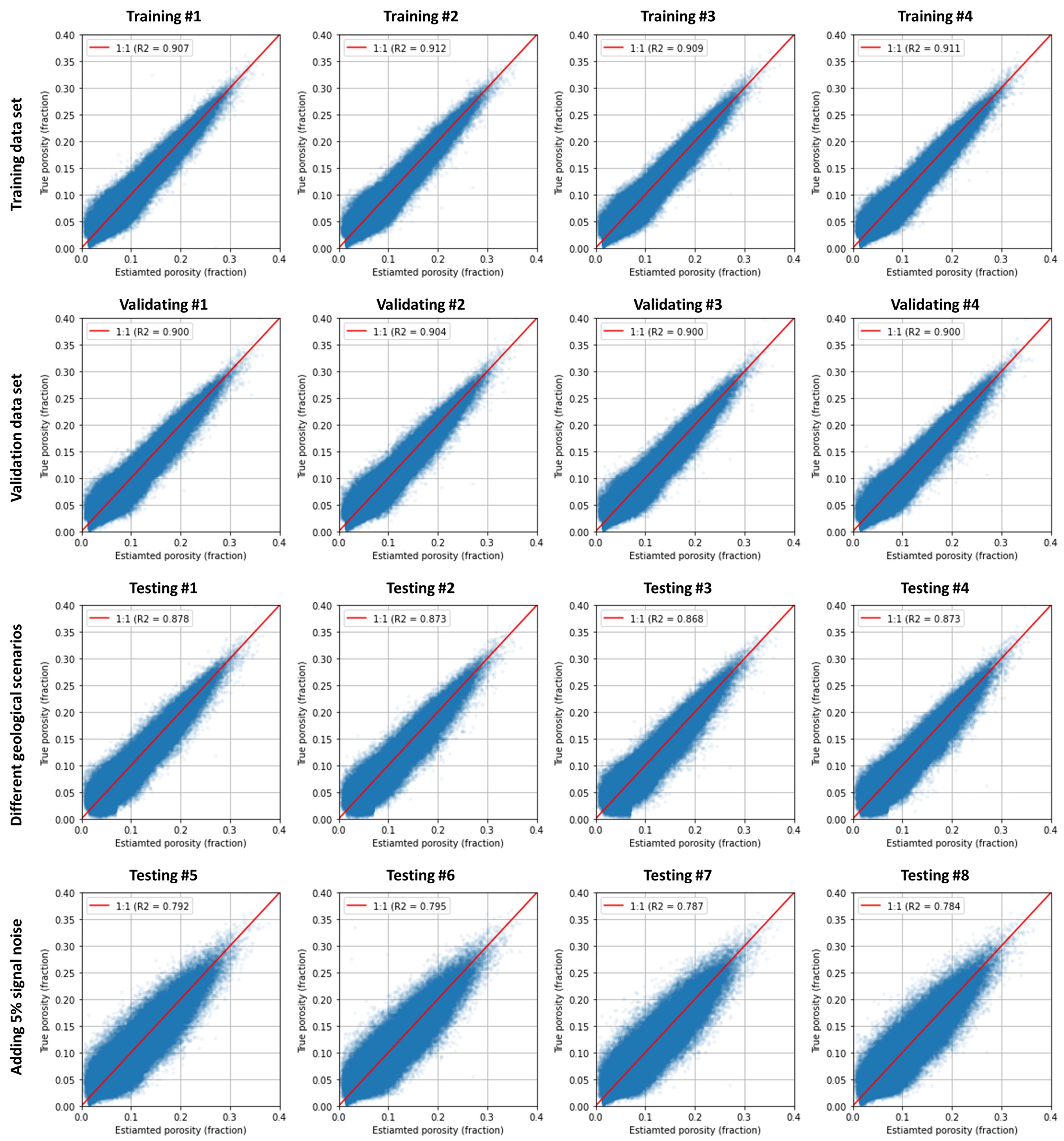}
\caption{Truth versus prediction plots of \texttt{ResUnet++}}
\label{fig:accuracy_plot}
\end{figure}

We visualize the truth versus prediction plots for training and validating data, as shown in the first two rows of Figure \ref{fig:accuracy_plot}. For both training and validating data set, most of the points along with 1:1 lines with over 0.9 coefficient of determination (i.e., R2 score). Besides, no significant bias (e.g., systemic under- or over-estimating) is observed in the plots. 

\subsection{Demonstration of robustness}
We perform the stress test to the trained \texttt{ResUNet++} to verify the expandability of applications to more practical cases, such as estimating porosity models for different geological scenarios or adding signal noise to seismic data. In addition, we conduct a comparative study using single frequency seismic data instead of three frequencies. 

\subsubsection{Application to different geological scenarios}
\begin{figure}[]
\centering
  \includegraphics[width=0.99\columnwidth]{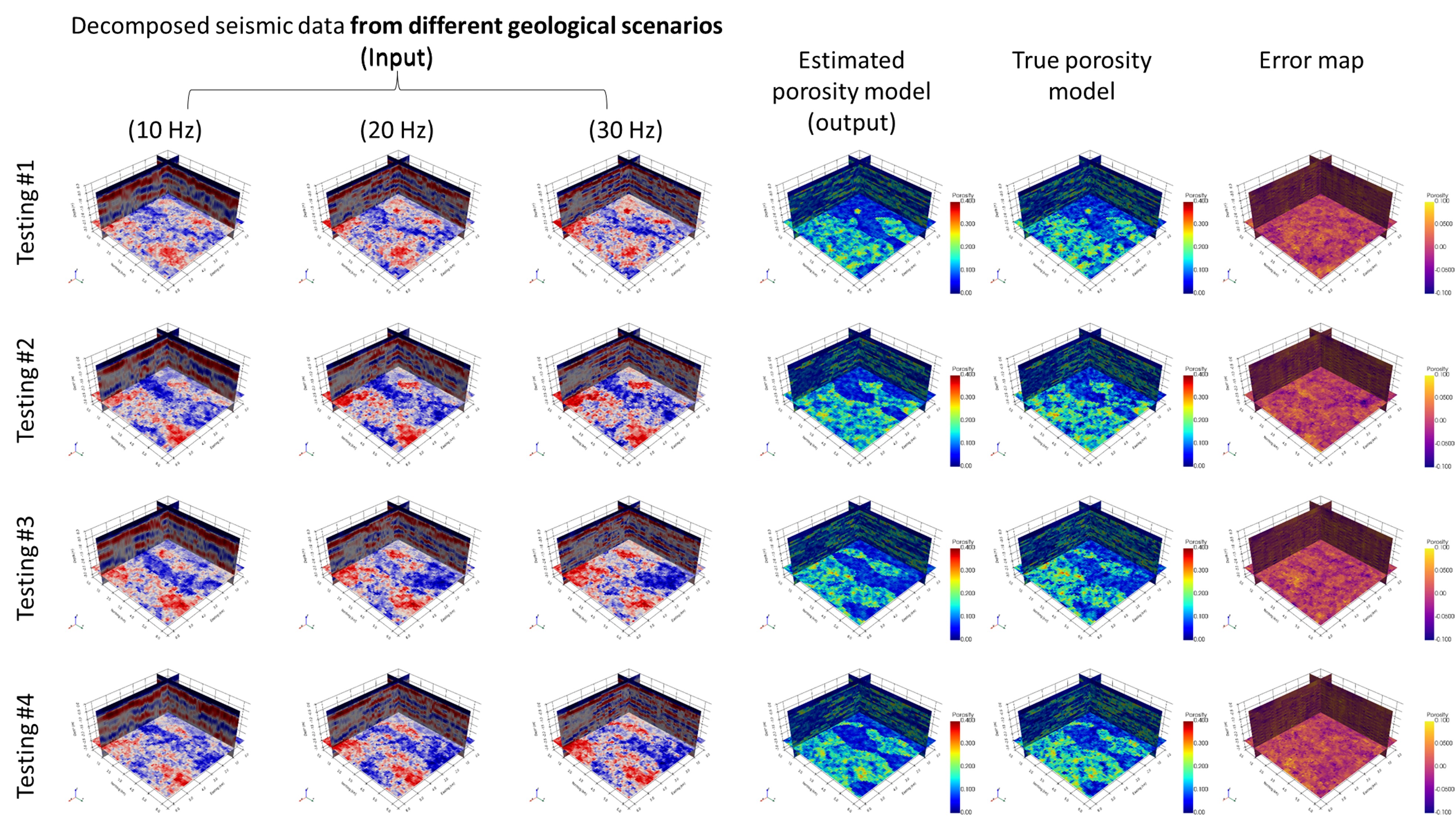}
\caption{Porosity estimation with different geological scenarios}
\label{fig:demo_geological}
\end{figure}

One of the sailent feature of the reference model and ensemble is that the major connectivity of rock facies is along with the easting direction, creating a larger high porosity connection along with the easting direction as shown in Figures \ref{fig:models} and \ref{fig:realizations}. We alter the major connectivity direction from easting to northing, which is the most significant change we can make, and generate another set of realizations as visualized in Figure \ref{fig:demo_geological}. Even though the \texttt{ResUNet++} has never seen this new configuration in training, the trained \texttt{ResUNet++} shows reliable prediction performance in Figure \ref{fig:demo_geological}. Most of the high porosity regions are well captured in the estimated porosity model, same as the truth models, and error maps have negligible values. 

The third row of Figure \ref{fig:accuracy_plot} shows the truth versus prediction of the ensemble with major connectivity directions. Same as training and validating data, most points are located along the 1:1 line, and R2 scores are over 0.86 in all cases. The degree of misfits between prediction and truth becomes slightly larger than the initial ensemble, but they are within an acceptable range. Therefore, with a properly trained \texttt{ResUNet++}, we can expand the applications to seismic data from different geological scenarios, such as different major directions and lengths. 

\subsubsection{Resilience against signal noise}
\begin{figure}[]
\centering
  \includegraphics[width=0.99\columnwidth]{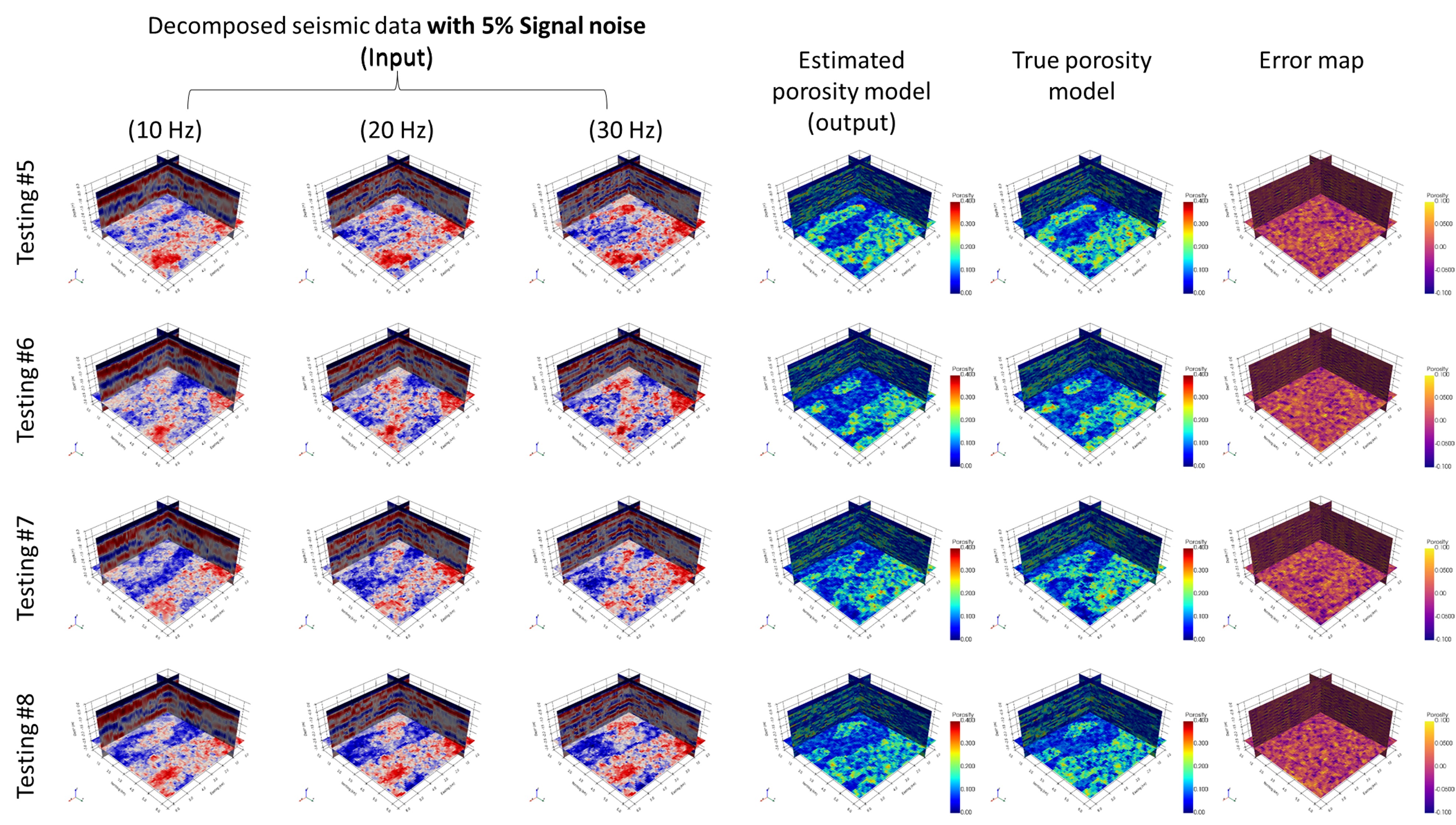}
\caption{Porosity estimation with 5\% signal noise}
\label{fig:demo_noise}
\end{figure}

After generating another set of realizations and calculating seismic data, we add 5\% signal noise to the seismic data (i.e., 5\% signal-to-noise ratio). Even though, none of these ensemble are used in training \texttt{ResUNet++}, Figure \ref{fig:demo_noise} shows that the trained \texttt{ResUNet++} successfully predict high and low permeability regions. The error maps show larger values than the demonstration of training and validating data set, but they are still close to zero without any systemic bias. The fourth row of Figure \ref{fig:accuracy_plot} presents the truth versus prediction of the ensemble with 5\% signal noise. The degree of misfits becomes larger than in previous cases, but R2 scores are still as high as 0.79. 

\subsection{Comparative study with single-frequency seismic data}
\begin{figure}[]
\centering
\subfigure[]{
  \includegraphics[width=0.85\columnwidth]{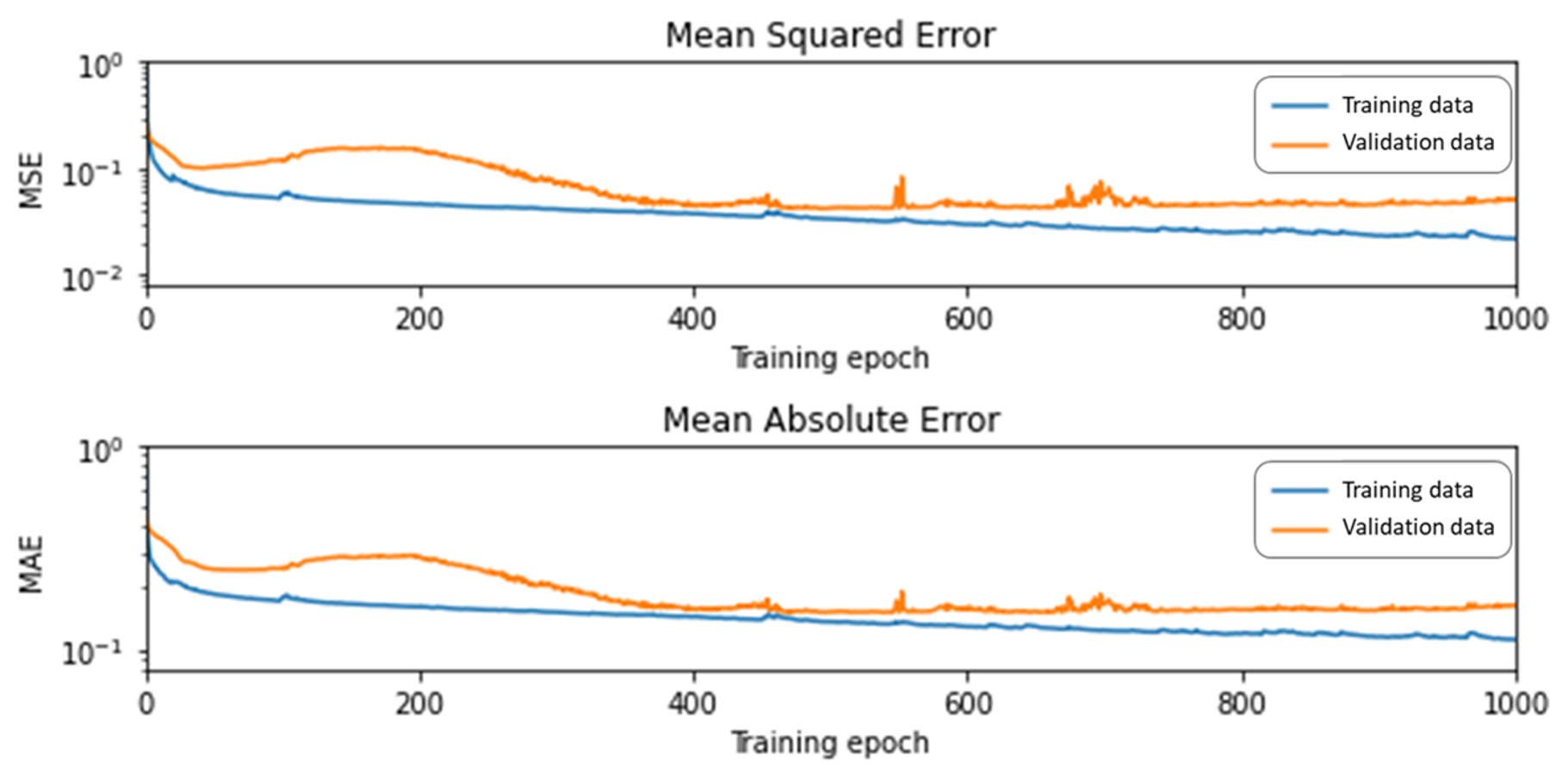}
  \label{fig:curve2}}
\subfigure[]{
  \includegraphics[width=0.85\columnwidth]{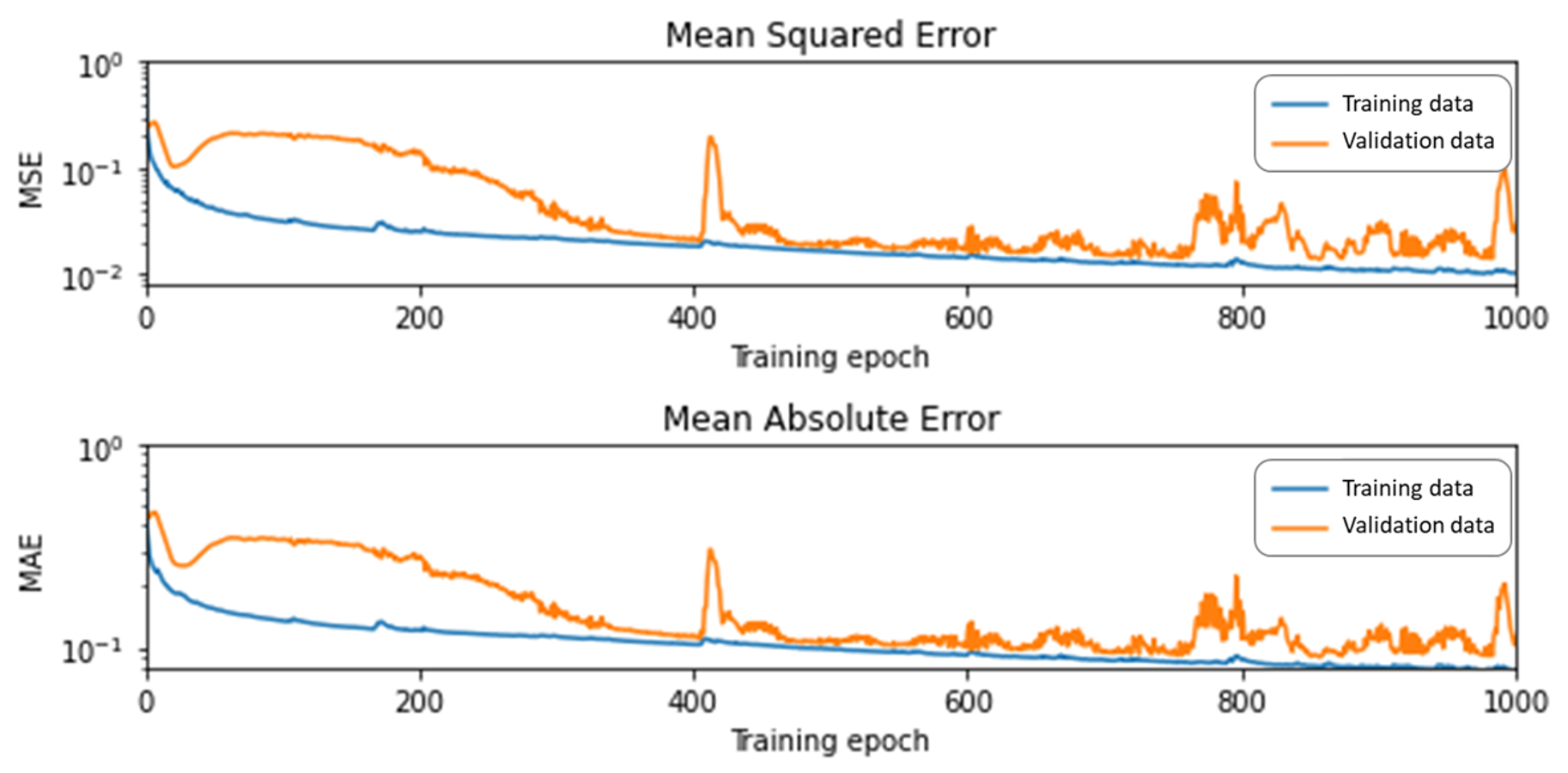}
  \label{fig:curve3}}
\caption{Loss curves of the ML-based models with (a) low- and (b) high-resolution seismic data only.}
\label{fig:err_curve}
\end{figure}

We perform a comparative study using single-channel seismic data. Instead of three different channels (e.g., low-, mid-, and high-resolution from 10~Hz, 20~Hz, and 30~Hz frequencies), we redesigned the first layer of \texttt{ResUNet++} to take a single frequency seismic data. We train the models by only taking either the lowest and highest resolution seismic data. Figure \ref{fig:err_curve} shows the loss curves of training and validating data. 

The model that only takes the lowest resolution seismic data results under-fitting as the losses of training and validating data do not effectively decrease until 1,000 epochs (Figure \ref{fig:curve2}). For example, the loss of training data is twice larger than the model with three different seismic data. Moreover, the loss of validating data becomes three times as large as the original model.

On the other hand, the model that only takes the highest resolution seismic data shows over-fitting. Figure \ref{fig:curve3} displays that the discrepancy between training and validating losses is significantly large at the end of the training, indicating the unreliable predictive performance of \texttt{ResUNet++} out of training data. Under- and over-fitting results with single seismic data show the importance of using multi-resolution seismic data to estimate porosity. 

\section{Conclusion}
An \texttt{ResUNet++}-based model is suggested to estimate porosity from three post-stack seismic volumes with different frequencies and demonstrated in the 3D channelized reservoirs. The suggested workflow directly takes the spectral decomposed seismic data and estimates the porosity model in less than one second. Moreover, we expand the application of the workflow to different geological scenarios and noisy seismic data, proving the robustness of the workflow. Even with Seismic data from different major connectivity orientations, the suggested method shows reliable prediction performance. The suggested workflow also stable prediction outcomes up to 5\% signal noise in seismic data, which is crucial for practical implementation for real data in future studies. Our work leads to the following three main conclusions:

First, by reversing the sequence of porosity estimation, the suggested workflow can shorten computational cost and alleviate subjectivity in interpretations in converting the acoustic impedance model from seismic data. The suggested workflow generates stochastic porosity realizations and converts  them to post-stack seismic volume, which is more straightforward than converting seismic data to acoustic impedance. Moreover, our results suggest that 100 realizations are enough to train \texttt{ResUNet++} in estimating porosity models of 64$\times$64$\times$32 grid, computationally available in typical desktop computers. 

Second, the \texttt{ResUNet++}-based model can linearity assumption between porosity and acoustic impedance, mitigating possible artifacts such as a poor representation of the impedance-porosity relationship and overestimation of porosity connectivity. Instead of relying on bi-variate correlation (e.g., co-kriging), the \texttt{ResUNet++}-based model enables to use spacial-correlations between seismic data and porosity in estimating porosity.

Third, using multiple spectral decomposed seismic data helps stabilize the training process of \texttt{ResUNet++}.  With the comparative study between single-frequency and multiple-frequency seismic data,  our results support the importance of employing multiple seismic data with different resolutions to estimate porosity.  Moreover, whereas conventional methods have no straightforward way to integrate multi-spectral decomposed seismic data, the \texttt{ResUNet++}-based model can easily take a set of seismic data with various frequencies thanks to the flexibility of neural networks.


\section*{Data availability statement}
The synthetic model and the raw/processed data required to reproduce these findings cannot be shared since authors do not have legal right to release the data.

\bibliographystyle{seg}
\bibliography{reference}

\end{document}